# Topological surface states induced by the magnetic proximity effect

Soichiro Fukuoka[1,+], Le Duc Anh[1,2,+,*], Masayuki Ishida[1], Tomoki Hotta[1], Takahiro Chiba[3,4], Yohei Kota[5], and Masaaki Tanaka[1,2,6,*]

[1] *Department of Electrical Engineering and Information Systems, The University of Tokyo, Japan*

[2] *Center for Spintronics Research Network (CSRN), The University of Tokyo, Japan*

[3] *Department of Information Science and Technology, Graduate School of Science and Engineering, Yamagata University, Japan*

[4] *Department of Applied Physics, Graduate School of Engineering, Tohoku University, Japan.*

[5] *National Institute of Technology, Fukushima College, Japan.*

[6] *Institute for Nano Quantum Information Electronics (NanoQuine), The University of Tokyo, Japan.*

+These authors contribute equally to this work.

*E-mail: anh@cryst.t.u-tokyo.ac.jp, masaaki@ee.t.u-tokyo.ac.jp

**The combination of magnetism and topological properties in one material platform is attracting significant attention due to the potential of realizing low power consumption and error-robust electronic devices. Common practice is to start from a topological material with band inversion and incorporates ferromagnetism via chemical doping or magnetic proximity effect (MPE). In this work, we show that a topological material is not necessary and that both ferromagnetism and band inversion can be established simultaneously in a trivial insulating material by MPE from a neighbouring ferromagnetic layer. This novel route is demonstrated using quantum transport measurements and first-principles calculations in a heterostructure consisting of 5-nm-thick $FeO_x$/1 monolayer of FeAs/3-nm-thick α-Sn. The Shubnikov–de Haas oscillations show that there is linear band dispersion with high mobility in the heterostructure even though a 3-nm-thick α-Sn single layer is a trivial semiconductor. Furthermore, first-principles calculations reveal that band inversion indeed occurs in this heterostructure, suggesting that the observed linear band is a topological surface state within this inverted gap. This work significantly expands the foundation for realizing magnetic topological materials in a myriad of trivial narrow-gap semiconductors.**

Nontrivial topological materials such as topological insulators (TIs)[1] and Dirac semimetals (TDSs)[2,3] have linear-dispersion boundary states called topological surface states (TSSs) that are typically protected by time-reversal symmetry (TRS) or spatial inversion symmetry (SIS). These boundary states result from the topologically nontrivial band structure in the bulk, are robust against local disorder, and exhibit spin-momentum locking[4,5], making them promising for manipulating coherent spin transport and quantum information. From both fundamental and practical points of view, an important issue is how this band topology interacts with a magnetic order that breaks the TRS, which would shed light on the exciting possibility of controlling these topological states.

Research on magnetic topological materials is rather scarce in comparison with a large body of nonmagnetic topological materials that have been predicted, fabricated, and characterized. One reason is that it is harder to calculate the band structure of such materials and more challenging to fabricate them[6]. The common approach to create magnetic topological materials[3,6] is usually to start with a non-magnetic topological material and to add magnetism by doping magnetic impurities or interfacing with a ferromagnetic layer[7–10]. However, TSSs may be eliminated due to the breaking of TRS[5], and the presence of a large amount of magnetic impurities causes severe degradation of the crystal structure and may alter the band topology of the host material. A theoretically proposed alternative route is to induce nontrivial phases from trivial band structures by inducing band inversion through exchange splitting[10]. By introducing ferromagnetic order in narrow-gap semiconductors, a magnetic exchange field inverts the valence and conduction bands if the exchange splitting is larger than the band gap[11]. If feasible, this approach would provide a much wider foundation for the realization of magnetic topological materials, using a myriad of trivial narrow-gap semiconductors. The transition from a trivial phase to a quantum anomalous Hall phase has been demonstrated to be effectively facilitated in $(Bi,Sb)_2Te_3$ through Cr doping, but only when a large external magnetic field of 5

T is applied [12]. These findings underscore the growing feasibility of inducing topological phase transitions via ferromagnetism.

In this work, we present clear evidence of a topological phase transition in a narrow-gap semiconductor by interfacing with a ferromagnet. To avoid the problem associated with conventional chemical doping of magnetic impurities, we approached this issue using the magnetic proximity effect (MPE) in a heterostructure consisting of a narrow-gap semiconductor and a ferromagnet. As illustrated in Fig. 1a, spin-splitting energy induced by MPE from the ferromagnet combined with spin-orbit coupling (SOC) of the narrow-gap semiconductor results in band inversion and the appearance of a TSS. This approach can prevent the degradation of carrier mobility and the formation of impurity bands in the host band structure. Therefore, it is possible to induce ferromagnetism and exchange splitting in the system while preserving the topological nontrivial phase with high mobility of carriers, which is promising as a platform for magnetic topological materials.

We used diamond-structure-type α-Sn[13–21] as the narrow-gap semiconductor. Bulk α-Sn is a zero-gap semiconductor that exhibits an inverted band structure at the Γ point. α-Sn can exhibit multiple topological phases under different strains and film thicknesses and is thus generating significant interest as a promising platform for topological physics. Theoretical and experimental studies have confirmed that when α-Sn is epitaxially grown on InSb or CdTe, it experiences in-plane compressive strain and a phase transition to a Dirac semimetal (DSM)[13,18]. However, band inversion in α-Sn thin films grown on InSb (001) only occurs when the thickness $t_{\alpha\text{-Sn}}$ exceeds 3 nm. At $t_{\alpha\text{-Sn}} \leq 3$ nm, thin α-Sn is a trivial insulating phase with a small band gap[13]. Recently, research has also been reported on α-Sn combined with ferromagnetic materials such as heterostructures[19–22] and Fe-Sn nanocrystalline materials[23,24]. These investigations achieved highly efficient spin-charge current conversion and observed a large anomalous Hall effect, marking a critical step in understanding and harnessing the potential of spintronic applications. However, the influence of introducing magnetism to the topological phase of α-Sn remains unclear, and further investigations are strongly required. In this work, we examined a very thin α-Sn film (≤ 3 nm), which is a trivial ultra-narrow-gap semiconductor without band inversion[13] (Fig. 1b, left). By growing a ferromagnet/α-Sn heterostructure, exchange splitting induced by the MPE occurs in the band structure of α-Sn, leading to the band inversion (Fig. 1b, centre). Also, α-Sn has strong SOC, so topological phase transition and TSS occur in the heterostructure (Fig. 1b, right).

We grew a heterostructure consisting of $FeO_x$ (5 nm)/FeAs (1 monolayer (ML) ~ 0.3 nm)/α-Sn (3 nm) on an InSb (001) substrate by molecular beam epitaxy (MBE), as illustrated in Fig. 2a. As a reference sample, we also grew a 3-nm-thick α-Sn single layer on an InSb (001) substrate. In the heterostructure, we inserted a 1-ML-thick FeAs layer with a zinc-blende crystal structure (the same crystal structure as InSb and α-Sn), in which stable ferromagnetism was shown to be established[25]. This ensures not only the preservation of interfacial crystallinity, but also leads to the MPE from the 1-ML-thick FeAs in addition to the overgrown 5-nm-thick $FeO_x$ layer. The cross-sectional scanning transmission electron microscopy (STEM) lattice image in Fig. 2b indicates that the 3-nm-thick α-Sn layer has a high-quality diamond-type crystal structure. Furthermore, both the STEM image and energy-dispersive X-ray (EDX) mappings in Fig. 2b confirm the presence of a 1-ML-thick FeAs layer with a zinc-blende crystal structure on top of the α-Sn layer. The magnetic properties of this heterostructure were characterized by superconducting quantum interference device (SQUID) magnetometry. Fig. 2c shows the magnetization measured at 4 K under a magnetic field parallel to the film plane, which indicates ferromagnetic hysteresis behaviour. Therefore, we expect the $FeO_x$/FeAs layers to affect the neighbouring α-Sn layer as a ferromagnetic layer.

We first used magneto-transport measurements to characterize the band structure of the reference 3-nm (18-ML) α-Sn single layer, which does not include a ferromagnetic layer (Fig. 3a). As shown in Fig. 3b, the longitudinal resistance $R_{xx}$ exhibits clear Shubnikov–de Haas (SdH) oscillations, which appear at 2 K and persist up to 8 K. After extracting the oscillatory component of the conductance, fast Fourier transformation (FFT) revealed oscillations with a single frequency component (10.0 T) denoted as $F$, as shown in Fig. 3c. The oscillatory component $\Delta G$ can be fitted using the Lifshitz-Kosevich (LK) theory[26] and is expressed as:

$$\Delta G(B,T,F,m,\mu,\gamma) = G\,\frac{\left(\frac{2\pi^2 k_\mathrm{B}T}{\hbar\omega}\right)}{\sinh\left(\frac{2\pi^2 k_\mathrm{B}T}{\hbar\omega}\right)}\exp\left(-\frac{\pi}{q\mu B}\right)\cos\left(2\pi\left(\frac{F}{B}-\gamma+\delta\right)\right) \qquad (1)$$

where $G$ is the proportionality coefficient, $\hbar$ is the Dirac constant, $k_\mathrm{B}$ is the Boltzmann constant, $\mu$ is the quantum mobility, and $m$ is the cyclotron mass. $\omega = qB/m$ is the cyclotron angular frequency, where $q$ is the elementary charge, $\gamma$ is a phase shift determined by the linearity of the band dispersion ($\gamma$ is 0 or 1 for a linear dispersion and 0.5 for a quadratic one), and $\delta$ is a phase constant that depends on the dimensionality ($\delta = \pm 1/8$ and 0 in a three-dimensional (3D) Fermi surface and a quasi-two-dimensional (2D) cylindrical Fermi surface, respectively). By fitting the temperature damping factor to the peak intensity of the Fourier-transformed spectra (mass fit), the effective mass was estimated to be $m$ = 0.109 $m_0$ (Fig. 3c inset). Using the obtained frequency and mass, we performed fitting using Equation (1) for the oscillatory component measured at $T$ = 2 K with $G$, $\mu$, and $\gamma$ as the fitting parameters (Fig. 3d). As a result, the following values were obtained: $\mu$ = 2,820 $cm^2V^{-1}s^{-1}$ and $\gamma$ = 0.97. These values were confirmed by a Dingle plot using the Dingle reduction factor, as well as a fan diagram based on the oscillatory factor, which both yielded similar results (see Supplementary Note 1). These analyses suggest the presence of a linear band at the Fermi level in the 3-nm-thick α-Sn layer.

Figure 3e shows the band structure of 3-nm (18- ML) α-Sn without a ferromagnetic layer obtained from our first-principles calculations. As designed, it presents an ultra-narrow-gap trivial state, which is just before the occurrence of band inversion (which occurs if the thickness is increased further). Although band inversion is lacking, both the conduction band and valance band are close to linear dispersion, which explains the experimental value of $\gamma$ =

0.97. From the effective mass and the mobility estimated experimentally, the Fermi level is expected to cross the heavy hole (HH) band (see Supplementary Note 2), which is consistent with the trend of thin α-Sn film tending to exhibit p-type carriers[27–29]. Therefore, we conclude that the 3-nm (18-ML) α-Sn is a trivial ultra-narrow-gap semiconductor.

Next, we conducted the same measurements and analyses for the ferromagnet (FM)/α-Sn heterostructure shown in Fig. 2a. As shown in Fig. 4a, strong SdH oscillations were also observed at 2 to 10 K. In contrast to the results of the single-layer α-Sn reference sample, the FFT-spectra in Fig. 4b reveal two oscillatory components ($F_{\mathrm{Low}}$ = 12.3 T, $F_{\mathrm{High}}$ = 33.4 T). We estimated their masses as $m_{\mathrm{Low}} = 0.165\ m_0$ and $m_{\mathrm{High}} = 0.168\ m_0$, respectively, which are greater than that of the reference sample. We calculated the SdH oscillation by the summation of two terms using Equation (1) and fit the result to the experimental data at 2 K using $G$, $\mu$, and $\gamma$ as the fitting parameters for both the $F_{\mathrm{Low}}$ and $F_{\mathrm{High}}$ components (Fig. 4c). The following values are obtained: $\mu_{\mathrm{Low}}$ = 28,900 cm$^2$V$^{-1}$s$^{-1}$, $\gamma_{\mathrm{Low}}$ = 0.28, $\mu_{\mathrm{High}}$ = 1,040 cm$^2$V$^{-1}$s$^{-1}$, and $\gamma_{\mathrm{High}}$ = 0.42. The values of $F_{\mathrm{Low}}$ were confirmed by Dingle and fan diagram plots, which both yielded similar results (see Supplementary Note 3). Interestingly, the result indicates that the $F_{\mathrm{Low}}$ band has remarkably high mobility comparable to that of the TSS observed in a topological Dirac semimetal α-Sn (thickness 9 nm) in our previous study[13]. Furthermore, the phase shift $\gamma_{\mathrm{Low}}$ = 0.28 exhibits a relatively linear dispersion in this band component. In contrast, the mobility of the $F_{\mathrm{High}}$ band is of the same order as that of the 2D heavy-hole (HH) band reported for α-Sn[13]. Additionally, the phase shift $\gamma_{\mathrm{High}}$ = 0.42 of the $F_{\mathrm{High}}$ band indicates a quadratic dispersion shape. These results are distinctly different from those of the 3-nm-thick α-Sn reference sample.

Then we examined the angular dependence of the SdH oscillations when rotating the magnetic field $\boldsymbol{B}$ from the perpendicular [001] direction to the in-plane direction, where $\boldsymbol{B}$ is parallel to the current $\boldsymbol{I}$ ($\boldsymbol{B}$ // $\boldsymbol{I}$; see Fig. 4d). $\theta$ is defined as the angle of $\boldsymbol{B}$ with respect to the [001] direction. As shown in Fig. 4e, the frequencies of the $F_{\mathrm{Low}}$ and $F_{\mathrm{High}}$ bands vary with $1/\cos\theta$ (Fig. 3h), indicating that both are 2D band components. In this heterostructure sample, the strong quantum oscillations can be attributed to only the α-Sn layer, suggesting that the $F_{\mathrm{Low}}$ and $F_{\mathrm{High}}$ bands correspond to the α-Sn layer. This conclusion is supported by the fact that the 2D conduction rules out the possibility of conduction derived from the underlying 100-nm-thick InSb layer. Furthermore, the $FeO_x$ layer is polycrystalline and is not likely to exhibit the quantum oscillations and large magnitude of magnetoresistance (MR) observed in Fig. 4a. Additionally, the FeAs layer is only 1-ML thick, reinforcing the dominance of α-Sn in the conductive behaviour of the system. These factors collectively indicate that the electrical transport properties detected in this heterostructure are predominantly characteristics of the α-Sn layer. The $F_{\mathrm{Low}}$ band is a linear 2D band with ultrahigh mobility, and from these results, it is expected to be a robust topological surface band resulting from a topological phase transition. There is also another oscillatory component, $F_{\mathrm{In\text{-}plane}}$, which is independent of the magnetic field angle emerging when $\theta$ is beyond 45 degrees. A Dingle plot of the 90-degree data indicates a parabolic band shape (see Supplementary Note 4). This component was caused by the parallel conduction in the InSb layer. These estimated quantum mobilities are consistent with the critical magnetic field where the corresponding SdH oscillations begin and supports the validity of our analysis.

To elucidate the effect of the ferromagnetic layer on the band structure of α-Sn, Fig. 4f shows the first-principles calculation results of the band dispersions of a 3-nm (18-ML)-thick α-Sn film with a top 1ML of FeAs. The s, p, and d orbital contributions are denoted by red, blue, and yellow, respectively. As expected from the mechanism explained earlier in Fig. 1, band inversion occurs around the Γ point due to the MPE, which would lead to the formation of a TSS to bridge this topological gap. The right panel of Fig. 4f shows the α-Sn slab in our calculations. The top surface is FeAs, and the bottom α-Sn surface is interfaced with a vacuum (corresponding to InSb in our experiment). Fig. 4g shows the surface band dispersions of the FM/α-Sn heterostructure under different directions of magnetization $\boldsymbol{M}$ of FeAs. Green and purple colours denote the contributions from the outermost Sn atom layers at the vacuum side and the FeAs side, respectively. The band components near the inverted gap at the Γ point are predominantly contributed by the α-Sn atoms at the vacuum surface (green). In contrast, the band components contributed by the Sn atoms at the interface with the FeAs layer (purple) largely hybridize with the d orbitals of Fe and are repulsed away from the Γ point. These results indicate that the TSS is formed at the vacuum surface but disappears at the FeAs/Sn interface. As shown in Fig. 4g, when $\boldsymbol{M}$ // [001] (left panel), a small exchange gap at –0.15 ~ –0.3 eV opens at the Dirac point of the TSS due to the TRS breaking[30]. A linear TSS is observable when $\boldsymbol{M}$ // $[\bar{1}10]$ (right panel), although it is shifted in the $k_x$ direction from the Γ point (see also Supplementary Fig. S4). This drastic change of the band structure of the α-Sn layer with rotation of the magnetization of the neighbouring FM layer is a direct consequence of spin-momentum locking and MPE, which possibly leads to great controllability of the topological states and physical properties in this heterostructure.

The TSS is expected to emerge around the Γ point between the inverted gap. To explain the two band components ($F_{Low}$ and $F_{High}$) observed in the SdH measurements, it follows that the Fermi level ($E_F$) is positioned at the pink dashed line in Fig. 4f and g. $F_{Low}$ corresponds to the TSS band (green curve in Fig. 4g, left panel), while $F_{High}$ corresponds to the bulk HH band (red flat band at the valence band top in Fig. 4f). This assignment is reasonable because a thin α-Sn film tends to exhibit p-type carriers[27–29], and there is good agreement between the experimental and calculated values of the distance between $E_F$ and the Dirac point and the Fermi wave number of these bands (see Supplementary Note 5). As revealed by our calculations, the HH band in the FM/α-Sn heterostructure (Fig. 4f) becomes flatter than that in the stand-alone α-Sn in the reference sample (Fig. 3e). On the other hand, the small exchange gap opening at the Dirac point of the TSS when $\boldsymbol{M}$ // [001] leads to a finite mass of the Dirac carriers. These changes of the HH and TSS in the FM/α-Sn heterostructure explain the relatively heavy masses ($m_{High}$ = 0.168 $m_0$ and $m_{Low}$ = 0.165 $m_0$, respectively) of the two bands and the slight deviation from linear dispersion ($\gamma_{High}$ = 0.28) of the TSS as estimated from the SdH analysis. The good agreement between our experiments and theoretical calculations confirms that a topological phase transition is indeed induced by MPE in our 3-nm α-Sn thin film interfaced with FM layers.

Finally, we report the observation of a rare odd-parity magnetoresistance (OMR) in our α-Sn samples, arising as a direct consequence of the MPE from the adjacent FM layers. As shown in Fig. 5a and b, when a magnetic field ***B*** is applied parallel to the current ***I***, the resulting resistance exhibits a pronounced antisymmetric behavior between positive and negative fields. Notably, Fig. 5b reveals a substantial odd-parity component of the magnetoresistance, which is linearly dependent on *B* and reaches approximately 2.5 Ω at 1 T —comparable to the zero-field resistance, corresponding to a remarkable ~100% change. Figure 5c presents the angular dependence of the OMR as ***B*** is rotated from the out-of-plane direction ($\boldsymbol{B} \perp \boldsymbol{I}$, $\theta = 0°$) to the in-plane direction ($\boldsymbol{B} \parallel \boldsymbol{I}$, $\theta = 90°$), where $\theta$ is defined as the angle between ***B*** and the [001] crystallographic axis of α-Sn. The OMR exhibits a complex angular pattern characterized by a four-fold symmetry along the [110] and [001] directions and a two-fold symmetry along $\theta \approx 41°$, the origin of which remains unclear at this stage (see Experimental Section for more detail). Importantly, this angular dependence is distinct from that of the Hall resistance (Fig. 5d), thereby ruling out any contribution from Hall effects to the observed OMR. Since OMR inherently requires time-reversal symmetry breaking, the exceptionally large OMR (~100% at 1 T) provides compelling evidence for the presence of MPE induced by the $FeO_x$/FeAs layers in α-Sn.

In summary, we have demonstrated transition from a topologically trivial phase to a nontrivial phase in α-Sn when it is interfaced with ferromagnetic layers. We grew a heterostructure consisting of a narrow-gap semiconductor α-Sn (3 nm), ferromagnetic FeAs (1 ML), and $FeO_x$ (5 nm) by MBE, and observed a linear 2D band with remarkably high mobility (28,900 $cm^2V^{-1}s^{-1}$) by quantum transport measurements. However, this band was not observed in the absence of the ferromagnetic layers. First-principles calculations revealed that the band inversion and formation of TSS occur in the α-Sn (3 nm) due to MPE and orbital hybridization with the neighbouring FeAs layer. Our work could pave a new way for the realization of magnetic topological materials from a broad spectrum of trivial narrow-gap semiconductors.

## Methods

*Sample growth*

We grew epitaxial α-Sn thin films with a diamond-type crystal structure (3 nm thick, equivalent to 18 monolayers (MLs)) on InSb (001) substrates at low temperature (−5°C) by molecular beam epitaxy (MBE). The MBE growth chamber was equipped with effusion cells for growing Sn and III−V semiconductors. To facilitate the formation of an atomically smooth α-Sn/InSb interface, a 100-nm-thick InSb buffer layer was first grown, followed by a 3-nm-thick α-Sn thin layer. In situ reflection high-energy electron diffraction (RHEED) with the electron beam azimuth along the $[\bar{1}10]$ axis showed streaky diffraction patterns, indicating a two-dimensional (2D) growth mode for both the diamond-type α-Sn and the zinc-blende-type InSb throughout the MBE growth. Prior to the growth of α-Sn, we formed an In-stabilized InSb surface, which exhibited c(8×2) reconstruction. The α-Sn layers grown on InSb experience in-plane compressive strain (~ –0.76%), as estimated in our previous study of α-Sn/InSb (001)[13]. This compressive strain and the thickness of 3 nm (18 MLs) drive the system into a trivial narrow-gap semiconductor phase. The reference sample was capped with $SnO_x$. For the heterostructure with $FeO_x$/FeAs, 1-ML-thick FeAs was grown on top of the α-Sn layer by depositing one atomic layer of As followed by Fe. As shown in the STEM lattice image of Fig. 2b, we observed the presence of 1-ML FeAs with a zinc-blende crystal structure at the interface. Next, a 2–3-nm-thick Fe layer was deposited, and the RHEED of the top surface of the Fe layer exhibited a ring pattern, indicating that it was deposited in a polycrystalline mode. The sample was then transferred to a sputtering chamber under a vacuum, where $SiO_2$ was deposited on top of the Fe layer. The top Fe layer was naturally oxidized and became $FeO_x$ after removal from the vacuum chamber, as indicated by the STEM and EDX results (Fig. 2b).

*Magneto-transport measurements*

The samples were patterned into Hall bars with dimensions of 50 μm x 200 μm by standard photolithography and Ar-ion milling. Since the α-Sn layer is very thin (3 nm), it has high thermal stability, and thermally induced phase transition to β-Sn during the process is prevented. Magneto-transport measurements were conducted in a temperature range of 2–10 K using a standard four-point probe method. A constant current $I$ of 100 μA was driven along the $[\bar{1}10]$ direction, while a magnetic field strength of up to 14 T was applied. The electrical conductance $G_{xx} = R_{xx}/(R_{xx}^2 + R_{yx}^2)$ was estimated from the longitudinal resistance $R_{xx}$ and the Hall resistance $R_{yx}$. The dependencies of $R_{xx}$ and $R_{yx}$ on the magnetic field direction were measured by rotating the sample in a fixed magnetic field $\boldsymbol{B}$ from the perpendicular [001] direction ($\boldsymbol{B} \perp \boldsymbol{I}$) to the in-plane $[\bar{1}10]$ direction ($\boldsymbol{B}$ // $\boldsymbol{I}$), as shown in Fig. 4d. We defined $\theta$ as the angle between the magnetic field $\boldsymbol{B}$ and the [001] direction.

*Analysis of SdH oscillations*

To analyze the SdH oscillations using LK theory, the oscillatory part, $\Delta G_{xx}$, was extracted from the raw experimental $G_{xx}$ data after removing the background component using polynomial fitting. Before the polynomial fitting, the experimental datasets of $G_{xx}$ were smoothened using a Savitzky–Golay filter to mitigate noise. They were then interpolated at a constant interval of $1/B = 5\times10^{-6}$ $T^{-1}$. This pre-processing step was done to circumvent the risk of introducing artificial weighting of the polynomial fitting of $G_{xx}$ as a function of $1/B$. Because the original data were acquired at uniform $B$ intervals, this led to disproportionately wide spacings between successive data points on the $1/B$ axis in the lower $B$ range. To improve the fitting accuracy, different polynomials were used for the data of different samples. A seventh-degree polynomial provided the best fit for the reference α-Sn single-layer sample, while a ninth-degree polynomial was utilized for fitting $G_{xx}$ in the $FeO_x$/FeAs/α-Sn heterostructure sample. After the oscillatory parts were extracted through the polynomial fitting, specific magnetic field ranges were selected as $1/B < 0.39$ $T^{-1}$ for the reference α-Sn single-layer sample and $1/B < 0.32$ $T^{-1}$ for the $FeO_x$/FeAs/α-Sn heterostructure sample. The Savitzky–Golay filter process was then performed on the data for further smoothening. This approach was done to eliminate artificial oscillations at the data edges caused by polynomial fitting and to avoid artificial peaks resulting from FFT. This procedure does not degrade the validity of the analysis.

We performed FFT on the data to determine the frequency of the oscillatory components. Before beginning, the data underwent pre-processing, where a Hann window was applied as the window function to recover periodicity and avoid artificial peaks near zero frequency in the FFT spectrum. A discrete window function was designed to have the same number of data points. Following this, arrays of "0" data with a finite length of 8 times the length of the original data were appended to both the start and the end of the data to increase the number of data points. This process was carried out to recover the resolution. Finally, the FFT was applied to the treated datasets.

The temperature dependence of each component's FFT peak intensity $A(B,T,m)$ in the oscillation parts was analyzed using the temperature-dependent factor of the LK theory, as specified in the following equation:

$$A(B,T,m) = G \frac{\left(\frac{2\pi^2 k_{\mathrm{B}} T}{\hbar\omega}\right)}{\sinh\left(\frac{2\pi^2 k_{\mathrm{B}} T}{\hbar\omega}\right)} \quad (2)$$

In this equation, the magnetic field $B$ is described as $1/B = (1/B_{\mathrm{max}} + 1/B_{\mathrm{min}})/2$, where $B_{\mathrm{max}}$ and $B_{\mathrm{min}}$ are the maximum and minimum values of the magnetic field range used in the FFT process. Through this analysis, the cyclotron masses for the components in the FFT spectra were determined. In the fitting process for estimating the mobility and phase, $\Delta G_{xx}$ at $T = 2$ K was modelled by a summation of oscillations as described by Equation (1) within the magnetic field range subjected to the FFT analysis. $G$, $\mu$, and $\gamma$ served as the fitting parameters, whereas parameters $F$ and $m$ were held constant to enhance the fitting accuracy.

Due to the time gap between measurements in Fig. 4a-c and Fig. 4d-e, there might be variations in sample quality, resulting in slightly different values for frequencies, etc. However, the trends of the obtained physical quantities remain consistent, which further supports the validity of the analysis and measurements.

*First-principles calculations*

First-principles calculations based on the density functional theory were performed using the Vienna Ab initio Simulation Package (VASP) with the projector augmented wave method[32,33]. Perdew–Burke–Ernzerhof (PBE)-type generalized gradient approximation (GGA) was adopted for the exchange-correlation functional [34]. Spin-orbit coupling was taken into consideration in the self-consistent calculations of the electronic structure. The cut-off energy of the plane wave basis was fixed to 500 eV in both bulk and slab geometry calculations. The Brillouin zone integrations were replaced by a sum over 12 × 12 × 12 and 8 × 8 × 1 Monkhorst-Pack k-point meshes. We adopted an α-Sn lattice constant of a = 6.4765 Å, which was determined by the total energy minimization of a bulk α-Sn system described by a cubic unit cell with the diamond-type crystal structure within the GGA+U method[35] (the on-site Coulomb potential U = –2.5 eV was used for this calculation, which is a very similar value to a previously reported one[36]). The obtained value of the lattice constant was in excellent agreement with the measured value, a = 6.4798 Å[37].

For both α-Sn and FeAs/α-Sn slab geometries, we used a layered tetragonal unit cell based on the diamond structure deposited along the [001] direction. The 18-ML slab geometry was adopted for the computation of the topological surface state (TSS) of a strained α-Sn single layer. Both edges of the slab were terminated with H atoms to remove dangling bonds. For the computation of a strained FeAs/α-Sn heterostructure, in the 20 ML slab geometry, the first 18-ML-thick slab was composed of Sn atoms, and the other 2 MLs were diamond-type FeAs consisting of one atomic layer of Fe and one atomic layer of As, as shown in the right panel of Fig. 4g. H-termination similar to the strained α-Sn case was used again. For both slab geometries, the surfaces were separated by a vacuum layer with a thickness of 1.8 nm. To calculate the strain effect on the topological electronic states, we considered a biaxial in-plane compressive strain of –0.76%[13]. Structural optimization for atomic position was performed with a fixed in-plane lattice constant. The correction energy of the on-site Coulomb repulsion, U = –2.5 eV, provides a better description of the electronic states in the bulk and slab geometries of α-Sn[13,36,38,39]. We confirmed that our GGA+U approach effectively reproduces the topological electronic structure around the band gap as well as the topological phase transition property obtained by means of similar GGA+U and LDA+U approaches[13,36,40].

*Fitting of the angular dependence of OMR*

As shown in Fig. 5c, the butterfly-shape angular dependence of the OMR on the magnetic field direction can be fitted using the following equation:

$$|OMR|\,(\Omega) = A_2 \cos 2(\theta - \theta_2) + A_4 \cos 4(\theta - \theta_4) + A_0. \quad (3)$$

Here, $A_0$, $A_2$, $A_4$ are the magnitude of the constant, two-fold symmetry, four-fold symmetry components, respectively. $\theta_2$, $\theta_4$ are the angles corresponding to the symmetry axes of the two-fold and four-fold symmetry components. From the fitting curve (blue curve) shown in Fig. 5c, we deduced $A_0$ = 1.75 ± 0.02 Ω, $A_2$ = 0.90 ± 0.03 Ω, $A_4$ = 0.39 ± 0.03 Ω, $\theta_2$ = 41 ± 1°, $\theta_4$ = 0°. These fitting results indicate that the OMR reaches local maximum when $\boldsymbol{B}$ // $\boldsymbol{I}$ or $\boldsymbol{B} \perp \boldsymbol{I}$, or when $\theta$ = 41 and 221° and local minimum when $\theta$ = 131 and 311°.

**Acknowledgements**

This work was partly supported by Grants-in-Aid for Scientific Research (19K21961, 20H05650, 23K17324, 24H00018), CREST program (JPMJCR1777) of Japan Science and Technology Agency, the UTEC-UTokyo FSI research granting program, and Spintronics Research Network of Japan (Spin-RNJ). A part of this work was supported by “Advanced

Research Infrastructure for Materials and Nanotechnology in Japan (ARIM)" of the Ministry of Education, Culture, Sports, Science and Technology (MEXT), Grant Number JPMXP1223UT0084.

**Funding:** This work was partly supported by Grants-in-Aid for Scientific Research (19K21961, 20H05650, 22K18293, 22K14591, 23K17324, 24H00018), CREST program (JPMJCR1777) of Japan Science and Technology Agency, and Spintronics Research Network of Japan (Spin-RNJ).

**Author contributions:** L. D. A. and M. T. supervised the study. S. F. and L. D. A. designed the experiments. S. F. and T. H. grew the samples. S. F. performed sample characterizations, transport properties and data analysis. T. C and Y. K performed first principles calculations. All authors discussed intensively and wrote the manuscript.

**Competing interests:** Authors declare no competing interests.

**Data and materials availability:** The data that support the findings of this study are available from the corresponding authors upon reasonable request.

## Figures

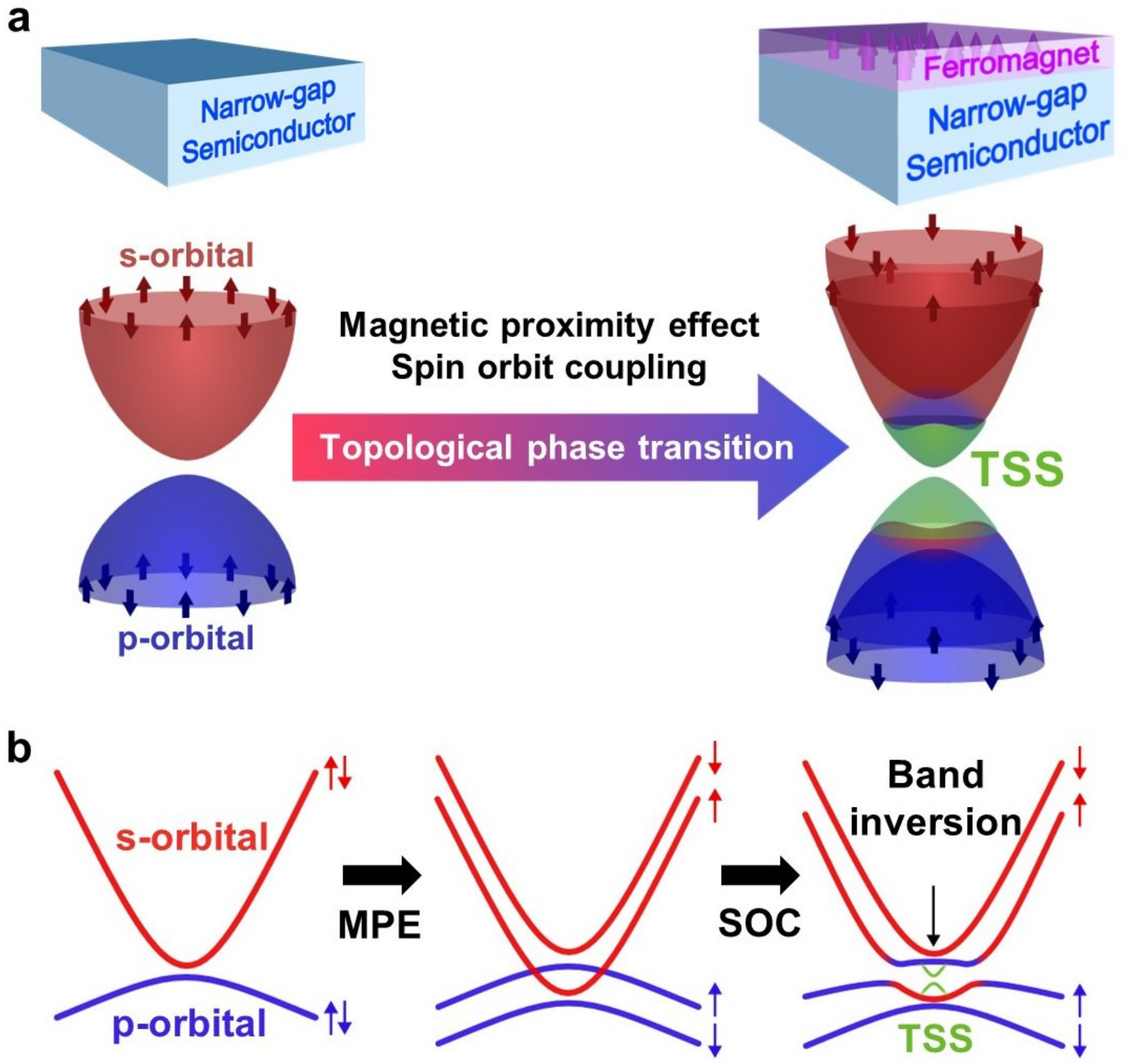

**Figure 1| Schematic illustration of a topological phase transition induced by the magnetic proximity effect (MPE).** (a) Left side is an electronic band structure of a narrow gap semiconductor (NGS). "↑" and "↓" represent spin up and spin down electrons. Right side is an evolution of the electronic band structure of NGS; a topological phase transition occurs by interfacing a NGS layer with a ferromagnetic layer. Here, s orbital, p orbital, and topological surface states (TSS) are denoted by red, blue and green colours, respectively. (b) Schematic evolution steps of the band structure of the narrow-gap semiconductor (NGS) under MPE and spin orbit coupling (SOC), resulting in an emergent topological phase. A trivial NGS without band inversion (left). In a ferromagnet / NGS heterostructure, exchange splitting induced by the MPE occurs in the band structure of NGS, leading to the band inversion (center). Because NGS has a strong SOC, topological phase transition occurs and a TSS emerges in the heterostructure (right).

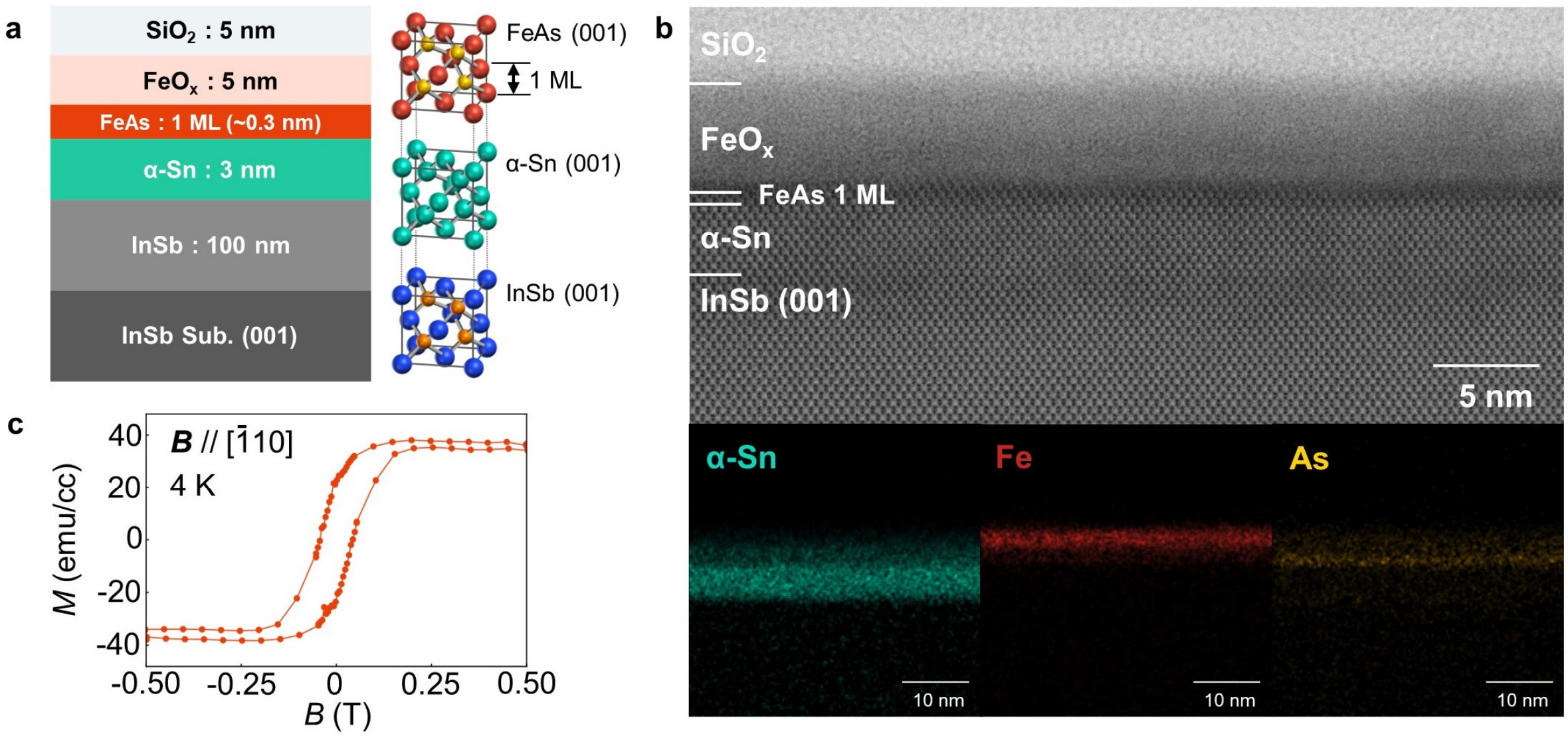

**Figure 2| Sample structure and properties.**(a) Left: Heterostructure consisting of $FeO_x$ / FeAs / α-Sn / InSb (001) examined in this study. Right: Crystal structures of InSb, α-Sn and FeAs, which have either zinc-blende or diamond-type structure. (b) Cross-sectional STEM lattice image and EDX elemental mapping images of the $FeO_x$ / FeAs / α-Sn / InSb (001) heterostructure. (c) Magnetic-field dependence of magnetization of the heterostructure, measured by SQUID magnetometry at 4 K, under a magnetic field parallel to the [$\bar{1}$10] axis in film plane.

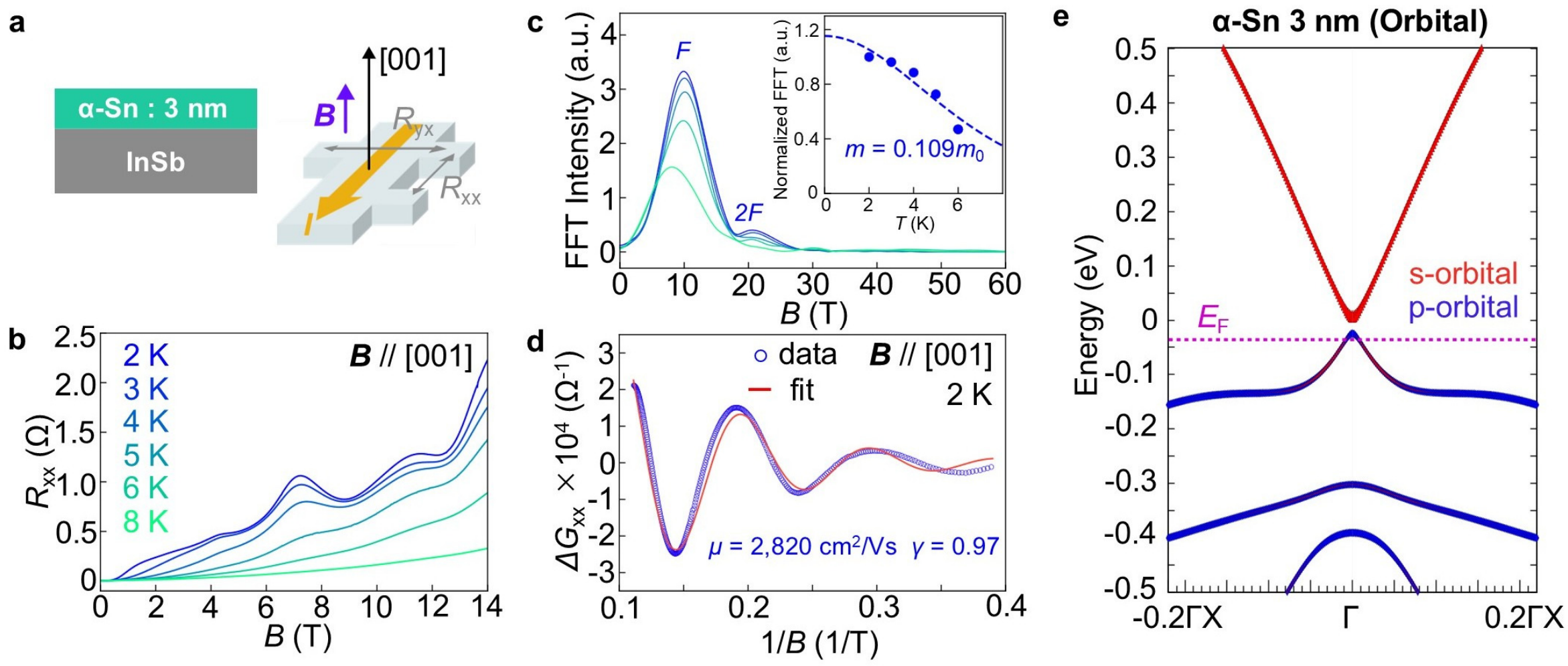

**Figure 3| Characterization of the band structure of a reference 3 nm-thick α-Sn single layer sample.** (a) Sample structure of the reference sample, a 3 nm-thick α-Sn single layer grown on an InSb (001) substrate, and magneto transport measurement setup. (b) $R_{xx} - B$ curves under a perpendicular magnetic field at various temperatures measured on the α-Sn reference sample. SdH oscillations can be clearly observed. (c) FFT spectra at various temperatures. The inset shows a fitting result of the temperature dependence of the FFT peak intensities, which gives the cyclotron mass value $m = 0.109m_0$. (d) SdH oscillation data at 2 K (blue circles) and a fitting curve based on the LK theory (red curve) of the oscillatory part $\Delta G_{xx}$, where there is a single frequency component $F$. (e) Calculated band structure of a 3 nm (18 ML)-thick α-Sn layer, in which the origin of the vertical axis corresponds to the calculated Fermi level. It exhibits a linear band shape but no band inversion, indicative of the electronic state before topological phase transition. The pink dashed line shows the Fermi level $E_F$ estimated from the analysis of the SdH oscillations.

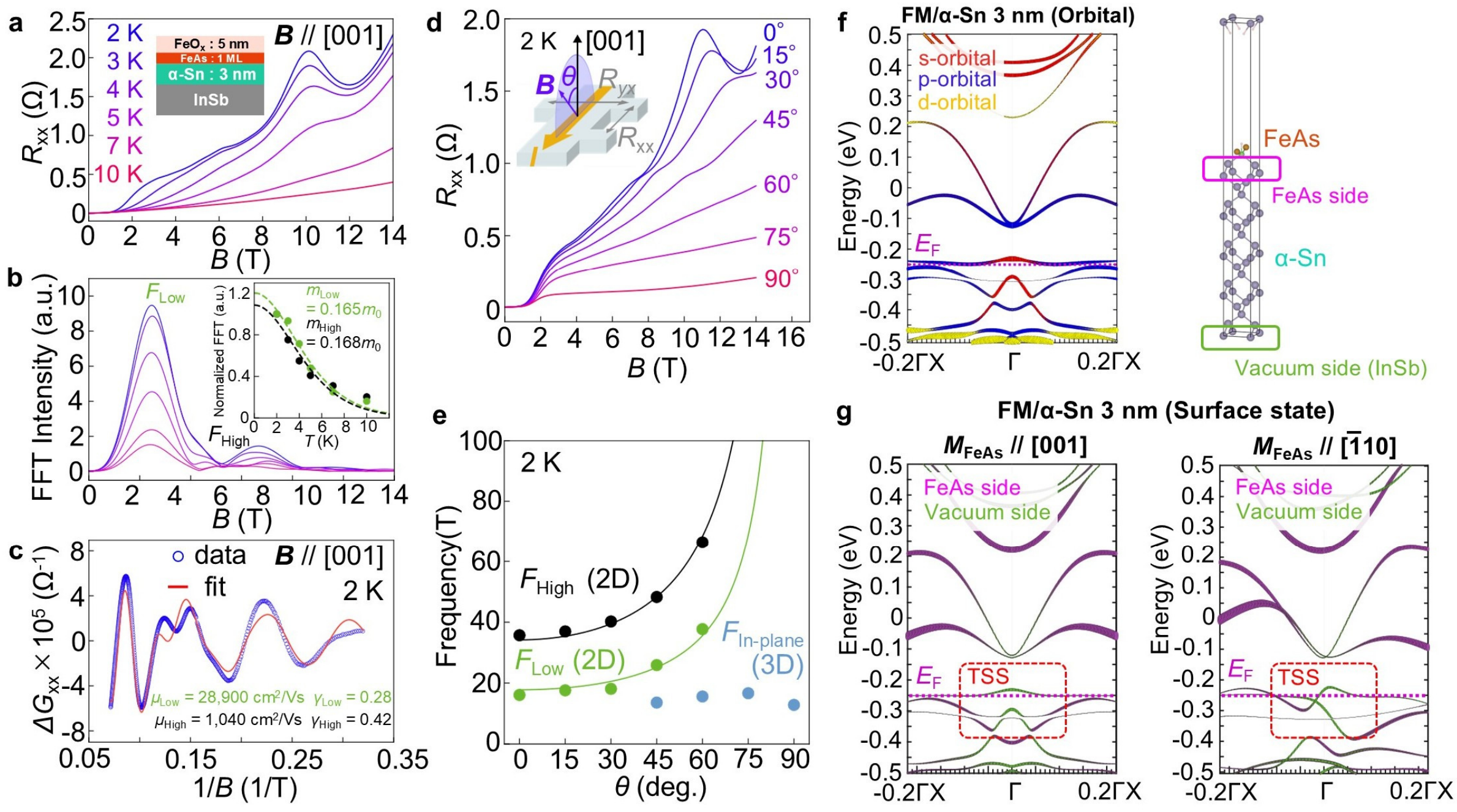

**Figure 4| Characterization of the band structure of the $FeO_x$/FeAs/α-Sn heterostructure.** (a) $R_{xx} - B$ curves under a perpendicular magnetic field at various temperatures from 2 – 10 K. (b) FFT spectra of the SdH oscillations at various temperatures. The inset shows fitting results of the temperature dependence of the FFT peak intensities, which give the cyclotron mass values $m_{\text{Low}}$ and $m_{\text{High}}$. (c) Oscillatory part $\Delta G_{xx}$ of the SdH data at 2 K (blue circles) and a fitting curve based on the summation of two oscillatory components $F_{\text{Low}}$ and $F_{\text{High}}$ (red curve). (d) Magnetic field angle dependence of $R_{xx} - B$. The magnetic field $\boldsymbol{B}$ is rotated from the vertical direction ($\boldsymbol{B} \perp \boldsymbol{I}$, $\theta = 0°$) to the current $\boldsymbol{I}$ direction ($\boldsymbol{B}$ // $\boldsymbol{I}$, $\theta = 90°$). $\theta$ is defined as the rotation angle of $B$ with respect to the [001] axis, as shown in the inset. (e) Angular dependence of the frequencies $F_{\text{High}}$, $F_{\text{Low}}$, and $F_{\text{In-plane}}$ (black, green, and blue gray circles, respectively) in the SdH oscillations. Experimental data are fitted by $a/\cos\theta$ (black and green curves), where $a$ is a proportionality coefficient. These results indicate the 2D characteristics of the band components corresponding to $F_{\text{High}}$ and $F_{\text{Low}}$. In contrast, $F_{\text{In-plane}}$ is independent of the magnetic field direction, indicating 3D characteristics. (f–g) First-principles calculations of the band structure of 1ML FeAs / 18 ML α-Sn, where the horizontal axis is the wavevector in the $x$ direction ($k_x$) along the Γ – X axis and the vertical axis is the electron energy. (f) Projection of the band component contributed from each type of orbital (s, p, d) is shown in different colours. Band inversion occurs at the Γ point due to the magnetic proximity effect of FeAs and spin-orbit coupling of α-Sn. The pink dashed line denotes the Fermi level $E_F$ (–0.25 eV) estimated from the analysis of the SdH oscillations. The right panel shows the slab in the calculation model[31]. (g) Projection of the band components contributed from the Sn atoms at the interfaces with the

top FeAs layer (purple) and the bottom vacuum (green) when the magnetization of FeAs (***M***) is in the [001] (left panel) and [$\bar{1}$10] (right panel) directions. The surface states at the vacuum side (corresponding to the InSb side in our experiment) are concentrated in the topological gap (surrounded by the red dashed curve). The TSS exhibits a small exchange gap due to the TRS breaking when ***M*** // [001], while it is shifted in the $k_x$ direction when ***M*** // [$\bar{1}$10] due to the spin-momentum locking effect and MPE (See Supplementary Fig. S4). The Fermi level $E_F$ intersects both the heavy hole (HH) band and the TSS.

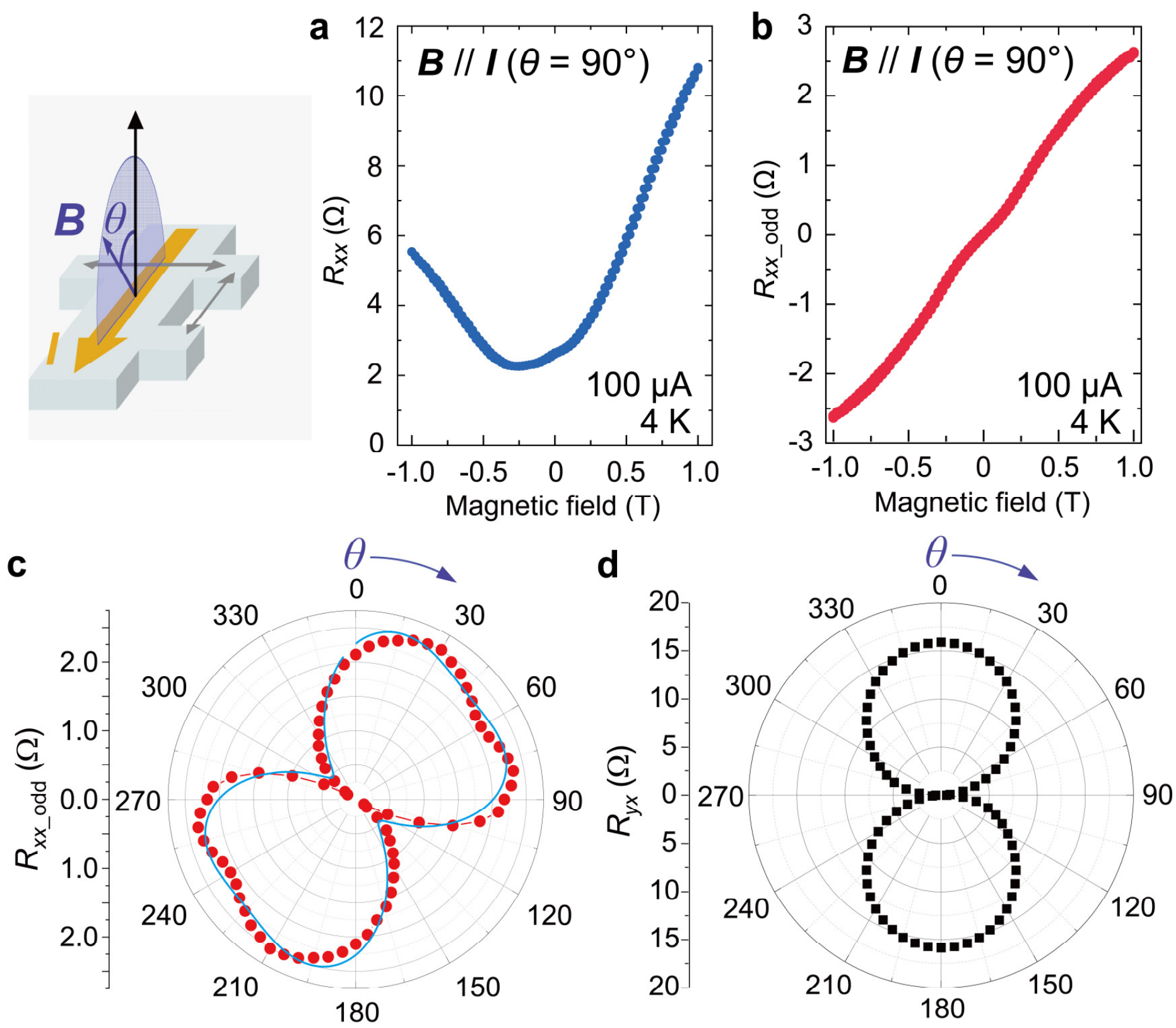

**Figure 5| Observation of odd-parity magnetoresistance (OMR).** (a) Magnetic field dependence of the longitudinal resistance $R_{xx}$ with a magnetic field ***B*** applied parallel to the current ***I*** (***B*** // ***I***). A prominent odd-parity component, characteristic of OMR, is observed. (b) Odd-parity magnetoresistance $R_{xx_odd}$, extracted from the antisymmetric part of the data in (a), showing a linear dependence on $B$. Angular dependence of the absolute values of (c) $R_{xx_odd}$ (left panel, experiment: red circles, fitting: blue curve) and (d) Hall resistance $R_{yx}$ (right panel) as a function of the magnetic field orientation, where the angle $\theta$ is defined as shown in the inset. All measurements were conducted at 4 K under a constant current of 100 μA.

# Supplementary Information

## Topological surface states induced by the magnetic proximity effect

*Soichiro Fukuoka, Le Duc Anh*, Masayuki Ishida, Tomoki Hotta, Takahiro Chiba, Yohei Kota, and Masaaki Tanaka**

### Supplementary Note 1. The Dingle and fan diagram plots of the reference sample

We independently determined $\mu$ and $\gamma$ in the reference sample using a standard Dingle plot with Equation (S1) and a fan plot with Equation (S2), respectively, as shown in Figure S1. The fan plot shows a phase shift of 0.98, which is consistent with the result of 0.97 in Figure. 3d. Additionally, the mobility was estimated to be 3,210 $cm^2V^{-1}s^{-1}$ from the Dingle plot, which also agrees with the result of 2,820 $cm^2V^{-1}s^{-1}$ in Figure. 3d.

$$\ln \frac{\Delta G}{\left(\frac{\left(\frac{2\pi^2 k_B T}{\hbar\omega}\right)}{\sinh\left(\frac{2\pi^2 k_B T}{\hbar\omega}\right)}\right)} = -\frac{\pi}{q\mu}\frac{1}{B} + C \tag{S1}$$

$$2\pi\left(\frac{F}{B} - \gamma\right) = (2N-1)\pi \tag{S2}$$

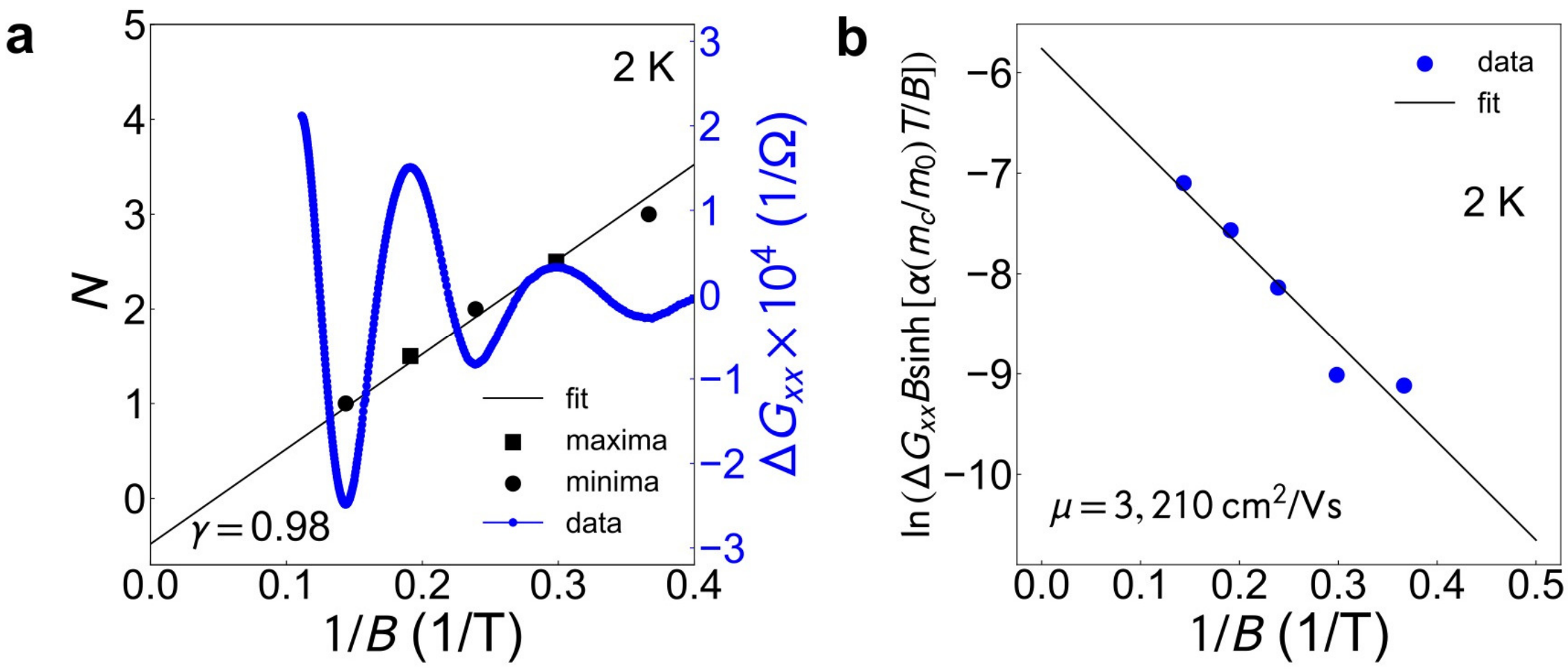

**Supplementary Figure S1.** (a) Fan plot of the component $F$. The blue line represents the oscillatory part. The minima (black circles) and maxima (black squares) are estimated from the fitting curve. By this fitting the phase $\gamma$ is estimated as 0.98. (b) Dingle plot of the component $F$, where $\alpha$ is $2\pi^2 k_B m_0/\hbar q$ ~ 14.7. By fitting the Equation (3) (black line) to the experimental data (blue circles), the quantum mobility $\mu$ is estimated to be 3,210 $cm^2V^{-1}s^{-1}$.

## Supplementary Note 2: Estimation of the Fermi level $E_F$ in the reference sample

Since the band corresponding to the comment $F$ exhibits linear dispersion, the distance from the Dirac point to the Fermi level $E_F$ takes the following value.

$$|E_F - E_{DP}| = \frac{2\hbar q}{m} F \ = 21.1\ meV \tag{S3}$$

Additionally, as the carrier type of α-Sn thin film tends to be p-type, we consider that the Fermi level is located at the position shown in Figure. 3e.

## Supplementary Note 3: Dingle and fan diagram plots of the $FeO_x$/FeAs/α-Sn heterostructure

We independently determined $\mu$ and $\gamma$ of the component $F_{Low}$ in the $FeO_x$/FeAs/α-Sn heterostructure using a standard Dingle plot with Equation (S1) and a fan plot with Equation (S2), respectively. To extract only the low-frequency band $F_{Low}$, measurements up to 9 T were performed. Based on the analysis shown in Figure. 4, where $\mu_{High} = 1{,}040$ $cm^2V^{-1}s^{-1}$, the condition for observing quantum oscillations

$$\mu B_c \sim 1 \tag{S4}$$

allows us to determine that the magnetic field $B_c$ at which quantum oscillations of the band begin to appear is about 10 T. In other words, it is expected that quantum oscillations of the $F_{High}$ band would not be observed in measurements up to 9 T.

As shown in Figure. S2, the fan plot shows a phase shift of $\gamma = 0.37$, which is consistent with the result of $\gamma = 0.28$ in Figure. 4c. Additionally, the mobility $\mu$ was estimated to be 16,000 $cm^2V^{-1}s^{-1}$ from the Dingle plot. This is also consistent with the fitting results using a summation of multiple components explained in the main manuscript (Figure. 4), demonstrating the reliability of the fitting approach. Note that due to the time gap between measurements in Figure. 4c and Figure. S2, there may be variations in sample quality, resulting in slightly different values from the fitting.

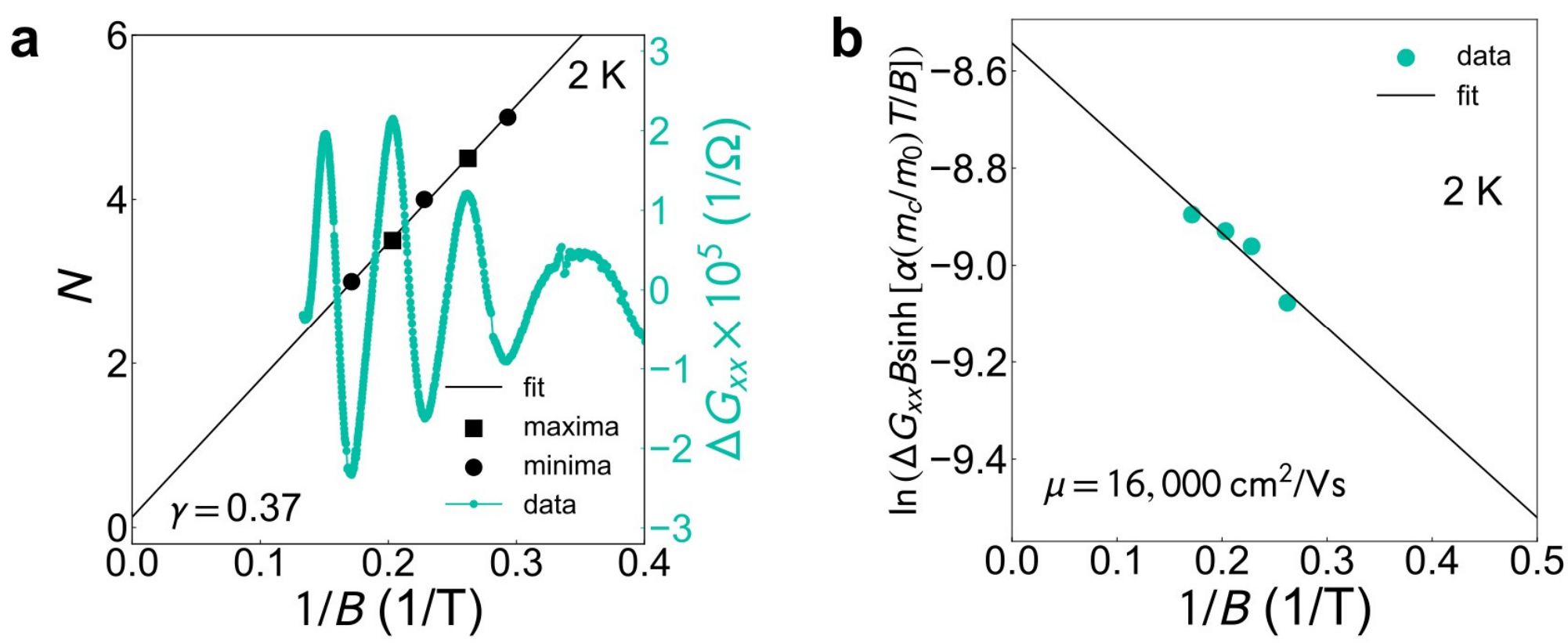

**Supplementary Figure S2**. (a) Fan plot of the component $F_{Low}$. The green curve represents the oscillatory part. The minima (black circles) and maxima (black squares) are estimated from the fitting curve. By this fitting, the phase $\gamma$ is estimated to be 0.37. (b) Dingle plot of the component $F$, where $\alpha$ is $2\pi^2 k_B m_0/\hbar q \sim 14.7$. By fitting the Equation (S1) (black line) to the experimental data (green circles), the quantum mobility $\mu$ is estimated to be 16,000 $cm^2V^{-1}s^{-1}$.

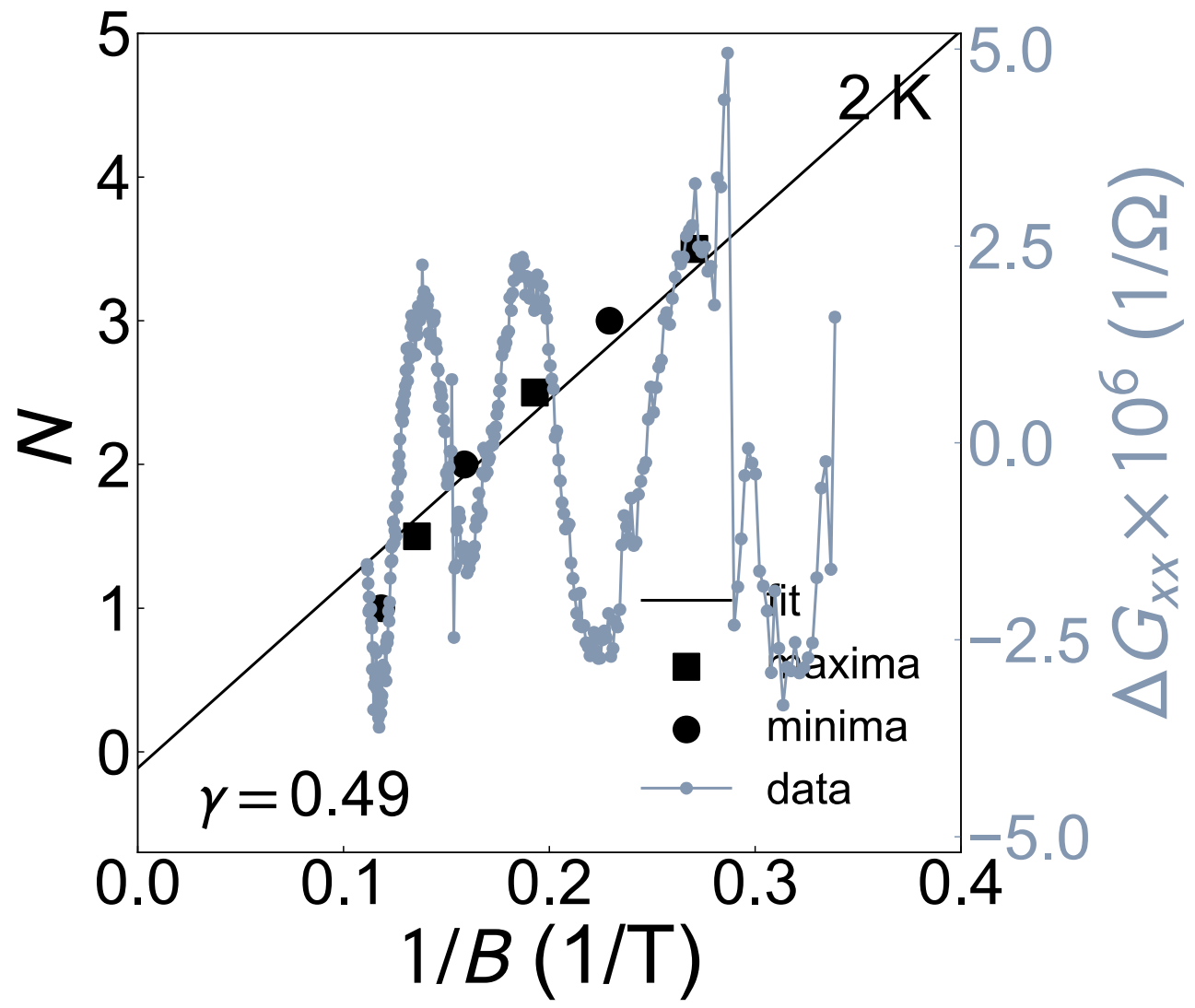

**Supplementary Figure S3.** Fan plot of the component $F_{\text{in-plane}}$. The grey line represents the oscillatory part. The minima (black circles) and maxima (black squares) are estimated from the fitting curve. By fitting Eq (S1) to the experimental data, the phase $\gamma$ is estimated to be 0.49.

### Supplementary Note 4: Fan diagram plot of the $F_{\text{In-plane}}$ peak in the $FeO_x$/FeAs/$\alpha$-Sn heterostructure with a magnetic field angle $\theta$ = 90 degree

As shown in Figure. S3, the phase shift $\gamma$ of the $F_{\text{In-plane}}$ component was found to be 0.49, indicating parabolic dispersion. This suggests that it originates from a trivial band typical of conventional electronic systems. Since it exhibits 3D conduction, the band component likely belongs to InSb. Because the extremal orbit was assumed to be a maximum, we used +1/8 as $\delta$ in Equation (1) of the main manuscript when estimating $\gamma$.

### Supplementary Note 5: Assignment of the bands in the $FeO_x$/FeAs/α-Sn heterostructure for the components $F_{\text{High}}$ and $F_{\text{Low}}$

From the results of band structure calculations, it is likely that the Fermi level crosses two bands, the top of HH band and the TSS, of the α-Sn layer. Based on this consideration, the wave number $k$ of $F_{\text{Low}}$ is $1.94 \times 10^6$ cm$^{-1}$ and $F_{\text{High}}$ is $3.19 \times 10^6$ cm$^{-1}$, thereby the area of the Fermi surface of $F_{\text{High}}$ is larger than that of $F_{\text{Low}}$. This leads to the assignments shown in Figure. 4 of the main text; $F_{\text{High}}$ and $F_{\text{Low}}$ correspond to HH and TSS, respectively. Furthermore, the high quantum mobility of $F_{\text{Low}}$ and the parabolic dispersion of $F_{\text{High}}$ also support the assignment.

Since $F_{\text{Low}}$ exhibits linear dispersion, we can estimate the distance from the Dirac point to the Fermi level to be 17.2 meV, by using equation (S3). The Fermi level $E_F$ was set from this value in Figures. 4f and g.

### Supplementary Note 6: Characterization of the DFT results by 2D massless Dirac Hamiltonian interacting to magnetic moments

Here we characterize the TSS band structure obtained by the DFT calculations based on 2D massless Dirac Hamiltonian which gives a simple model for the electronic structure of a TSS. As shown in Figure. 4f, we model the TSS on the bottom surface of Sn as 2D massless Dirac electrons that are exchange-coupled to a homogeneous magnetization $\boldsymbol{M}$ of FeAs via the magnetic proximity effect (MPE). Because of a very thin film of the Sn layer, we assume that MPE overcomes the slab system. Then, a low-energy effective Hamiltonian is given by [S1,S2]

$$\hat{H}_{2D} = -i\hbar v_F \hat{\boldsymbol{\sigma}} \cdot (\boldsymbol{\nabla} \times \hat{\boldsymbol{z}}) + \Delta \hat{\boldsymbol{\sigma}} \cdot \boldsymbol{M} \tag{S5}$$

where $\hbar$ is the reduced Planck constant, $v_F$ is the Fermi velocity of the Dirac electrons, $\hat{\sigma}$ is the Pauli matrix operator, and $\Delta$ is the MPE-induced exchange energy that is responsible for the hybridization between s,p and d orbitals. The Hamiltonian (S5) leads to the energy-momentum dispersion

$$E_{\mathbf{k}s} = s\sqrt{\left(\hbar v_F k_x + \Delta M_y\right)^2 + \left(\hbar v_F k_y - \Delta M_x\right)^2 + \left(\Delta M_z\right)^2} \tag{S6}$$

where $\hbar\boldsymbol{k}$ is the momentum measured relative to the Γ point of the 2D Brillouin zone and $s = \pm$ denotes the index of the upper (electron) and lower (hole) bands. The energy-momentum dispersion (S6) for fixed magnetization directions is schematically shown in Figure. S4 (top illustrations). As seen in Figure. S4 (bottom panels), the TSS band structure for the cases of ***M*** // [001], ***M*** // [$\bar{1}$10], ***M*** // [110] obtained by the DFT calculations are in excellent agreement with predictions by the Dirac electron model. Note that ***M*** // [$\bar{1}$10] and ***M*** // [110] directions correspond to $M_y$ and $M_x$ components in Eq. (S6), respectively.

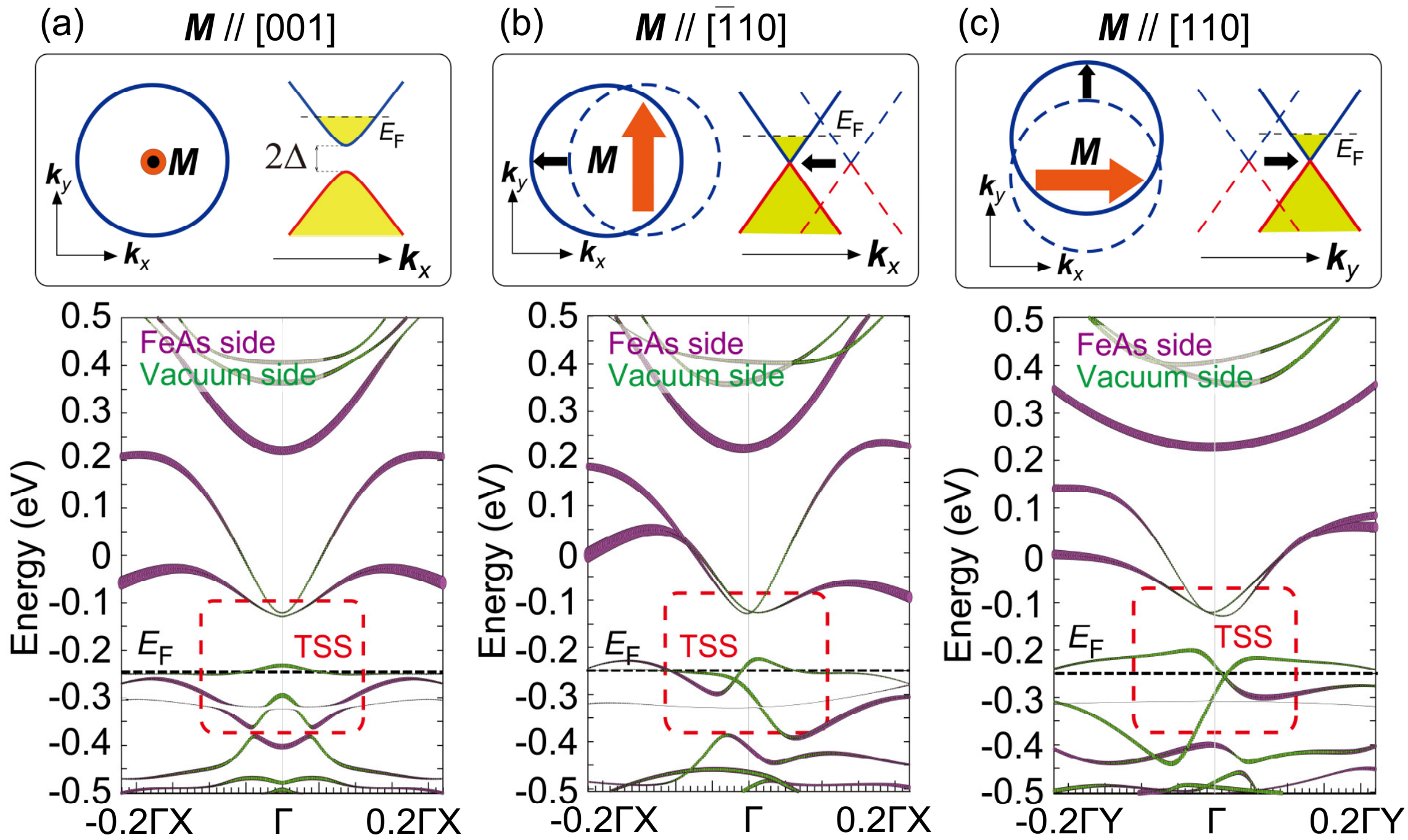

**Supplementary Figure S4.** Illustrations of the topological surface state (TSS) (top panels) and calculated band structures along the Γ – X ($k_x$) or Γ – Y ($k_y$) axis (bottom panels) are shown for the cases of (a) ***M*** // [001], (b) ***M*** // [$\bar{1}$10], (c) ***M*** // [110]. The bottom panels show the projection of the calculated band components of the FeAs/α-Sn heterostructure contributed from the Sn atoms at the interfaces with the top FeAs layer (purple) and the bottom vacuum surface (green) corresponding to the interface with InSb in our experiment. The vacuum surface states are concentrated in the topological gap (surrounded by the red dashed curve). The interface band structure of Sn changes drastically with the magnetisation direction (***M***) of FeAs due to spin-momentum locking. When ***M*** // [001], the TSS opens a small exchange gap of 2Δ at (–0.15 ~ –0.3 eV) due to time reversal symmetry breaking. A linear TSS is observable when ***M*** // [$\bar{1}$10] and ***M*** // [110] as shown in b) and c), respectively, which is shifted away from the Γ point due to the spin-momentum locking and MPE.

**Supplementary References**